# Frequency Observations and Statistic Analysis of Worldwide Main Power Grids Using FNET/GridEye

Xianda Deng[1], *Student Member, IEEE*, Hongyu Li[1], *Student Member, IEEE*, Wenpeng Yu[1], Wang Weikang[1], *Student Member, IEEE*, Yilu Liu[1, 2], *Fellow, IEEE*

[1.]Department of Electrical Engineering and Computer Science, The University of Tennessee, Knoxville, USA
[2.]Oak Ridge National Laboratory, Oak Ridge, USA
xdeng6@utk.edu, hongyu.li.1990@gmail.com, wyu10@utk.edu, wwang72@vols.utk.edu, liu@utk.edu

*Abstract*—**With the increasing renewable energy sources, concerns about how renewable energy sources impact frequency have risen. There are few reports regarding power frequency status in worldwide main power grids and what are differences of frequency status between power grids in mainland and island. FNET/GridEye, a wide-area measurement system collecting frequency and phase angle data at the distribution level, provides an opportunity to observe and study the power frequency in different power grids over the world. In this paper, 13 different power grids, spreading at different mainland and islands over the world, are observed and compared. A more detail statistical analysis was conducted for typical power grids in three different places, e.g., U.S Eastern Interconnection (EI), Egypt, and Japan. The probability functions of frequency based on the measured data are calculated. The distributions of frequency in different power grids fall into two categories, e.g., single-peak distribution and multi-peak distribution. Furthermore, a meaningful insight that the single-peak distributions of the frequency almost follow the normal distribution is found. The frequency observations and statistic analysis of worldwide main power grids using FNET/GridEye could help the power system operators understand the frequency statistical characteristic more deeply.**

*Index Terms*-- **Worldwide power grids, frequency observations, Statistic analysis, PMU application, big data, FNET/GridEye.**

## I. INTRODUCTION

Renewable energy sources have been promoted all over the world in recent years, due to its clean, low-cost and inexhaustible features, compared to the traditional power generation [1]-[2]. Meanwhile, the intermittent and unpredictable natures of renewable energy sources enhance the system frequency fluctuation, which brings challenges to power system operations and grid security [2]-[9]. Numerous factors determine power grid power frequency deviation [10], e.g. system capacity, load types, regulation requirements, power imbalance and etc. To maintain the system stability and meet system frequency regulation, more reserve power sources are required to accommodate the increasing renewable energy sources in the power grids. To compromise between stability risks and expensive cost of reserve power, research has been conducted to develop new control strategies and reserve sizing techniques [10]-[12]. In another direction, some researchers proposed that a flexible frequency operation strategy would satisfy load demands with limited reserve power sources [14]. It is noticed that although many studies have been contributed into moderating the problem, there are few observations and reports about what are the present power frequency status and frequency fluctuation ranges of worldwide power grids [15]-[17]. Power frequency deviation varies in different power grids, especially for the grids on small islands. Hence, it is worthwhile to investigate that power frequency status in worldwide major power grids and what are power frequency differences between the mainland and island.

Wide-area monitoring systems (WAMS), consisting of advanced measurement technology, information tools, and operational infrastructure, facilitates the understanding and management of the increasingly complex behavior exhibited by large power systems [18]. With the real-time, global positioning system (GPS) time-synchronized measurements at high data rates, WAMS reveals unprecedented insights into power grid dynamics and will be the next-generation operational-management systems [18]-[25]. However, the worldwide WAMS systems have not been built yet, let alone the frequency observations for the worldwide power systems.

FNET/GridEye, a kind of wide-area measurement system (WAMS), has been developed and operated over decades [23]. Hundreds of frequency disturbance recorder (FDR) are installed over the world, which is sending high time resolution measurements to FNET/GridEye servers. The FNET/GridEye servers are located in the University of Tennessee, Knoxville (UTK) and Oak Ridge National Laboratory (ORNL). With the valuable measurements, FNET/GridEye, as an independent observer, observes power grid operation frequency status in different regions over the worlds [23]. Based on measurements from FNET/GridEye, static analysis of the power system frequency and rate of change of frequency (ROCOF) in different power grid have been published and provided guidance in different research areas [15]-[17]. In this paper, the frequency data of worldwide major mainland and island power grids are analyzed and studied from the view of statistics, based on FNET/GridEye measurements.

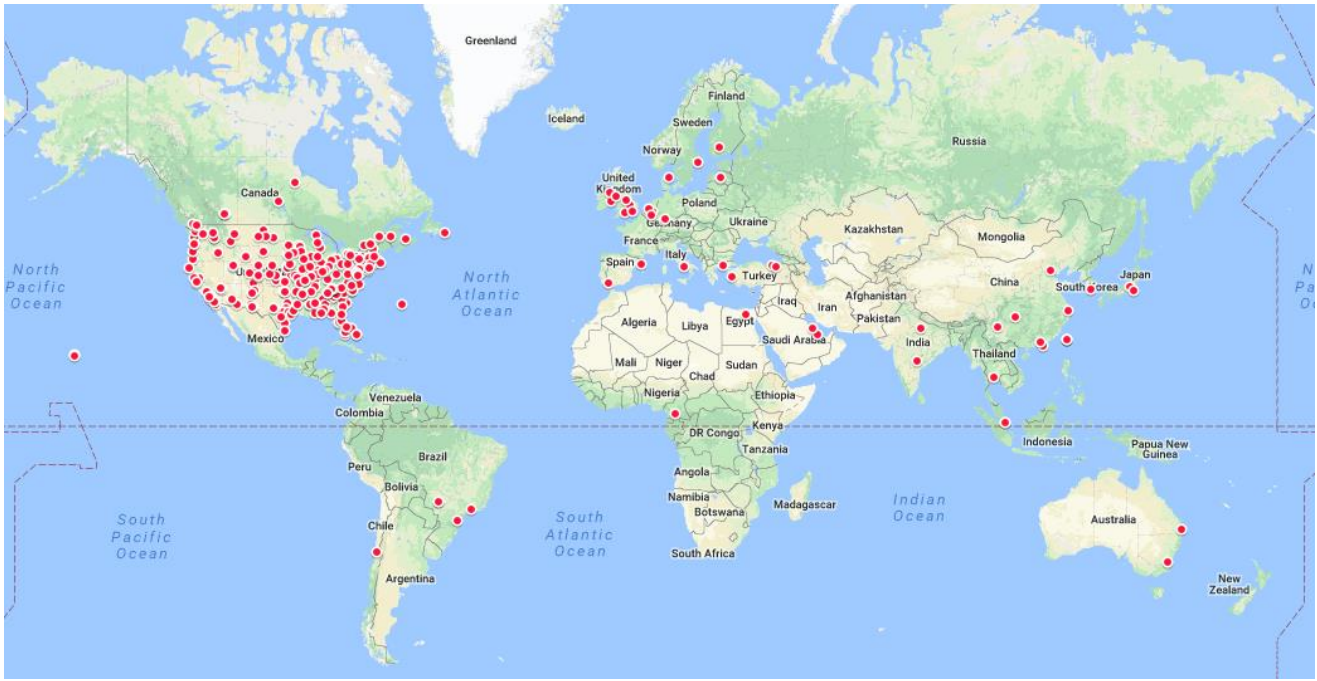

Figure 1. FDR Worldwide deployment map.

Though research has been conducted on studying the power frequency in the actual power grids, a few papers reported the power frequency in the worldwide main power grids, and thus the insights for the worldwide power system frequency are seldom offered. Using the data from FDRs, this paper offers the statistical analysis of the frequency in the different regions, which is helpful for the frequency study.

This paper is organized as follows: Section II provides a brief introduction to FNET/GridEye and FDR. Frequency observation and analysis of major mainland and island power grids worldwide are presented in Section III. In Section IV, a more detail statistical analysis of three typical types of power grids are given. The conclusion of the paper is drawn in section V.

## II. FNET/GRIDEYE OVERVIEW AND OBSERVATION DATA PREPROCESSING

FNET/GridEye is the first WAMS system ever designed to be deployed at the distribution level, whose mission is to pioneer and promote the WAMS technologies in electric power utilities [20], [23]. Data processing, visualizing and analyzing applications have been implemented on the FNET/GridEye system to process near real-time measurements, collected by variable types of FDRs [16]. Up to date, 297 FDRs are deployed in 31 counties across the world. The locations of FNET/GridEye FDR sensor over the world are shown in Fig. 1. FDR deployment map demonstrates that the FNET/GridEye could observe the frequencies of worldwide main power grids.

One of the most distinctive features of FNET/GridEye is the employment of low-cost and high-accuracy sensor, installed at the distribution level, which provides the probability to install FDR sensors over the world with low cost. After years of efforts, contributed by PowerIT group at UTK, three generations of FDR with different features have been developed, including smartphone-based FDR, magnetic and electric field based FDR, and universal grid analyzer (UGA) [26]-[28]. A photo of UGA is shown in Fig. 2. At present, the sample rate of the FDRs in FNET/GridEye is 10 times per second. The FDRs transmit the nearly real-time data with GPS synchronized timestamp to FNET/GridEye server via the internet.

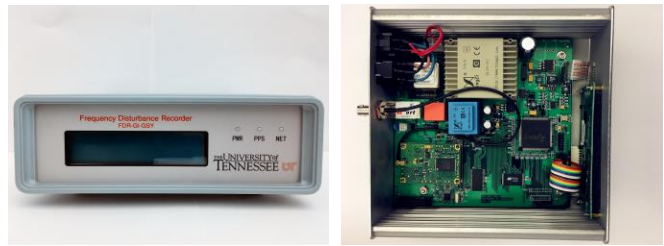

Figure 2. The prototype of UGA.

At the FNET/GridEye data center, a data concentrator is employed to receive and process FDR steaming data. After filtering out bad data, all the data are achieved into a database for data analysis [23]. The streaming data is visualized and published at http://fnetpublic.utk.edu/, while historical data is accessible with further privilege authorization. With a vast volume of frequency measurement data collecting over the world, FNET/GridEye provides an opportunity to observe and analyze different power grids frequency status. In the next, the frequency data collected by FDRs will be analyzed.

For a comprehensive observation purpose, 6 mainland power grids and 7 island power grids, which spread over America, Asia, Europe, Oceania, and Africa, are observed and analyzed in this section. According to [29], [30], frequency in different voltage levels is identical, while frequency differences in different regions of one system are relatively small. Measurements of one FDR in each country is selected for the analysis. The three-month frequency measurement data of the power grids collected by FNET/GridEye are retrieved from the database. The measurement data used in this studied reflect both normal operation and transient status of the power grids.

All the simulations are conducted on a computer running a 64-bit Windows 10, with a 3.60 GHz Intel I7-7700U CPU and 16 GB memory. The sample ratio of FDR is 10 points per seconds and the studied period are three months, thus the overall volumes of the 13 selected systems measurements data are 9.01 GB. To perform an efficient analysis, R language is employed to implement data processing and analyzing.

The nominal frequencies of the selected power grids include both 50 Hz and 60 Hz, which is shown in TABLE I. Hence, the frequency measurement data are converted to per unit value (i.e., the unit is p.u.) for a convenient comparing purpose.

TABLE I. THE OPERATION FREQUENCY OF POWER GRIDS WORLDWIDE

| Country | Continent | Nominal Frequency (Hz) |
|---|---|---|
| EI, U.S. | America | 60 |
| WECC, U.S. | America | 60 |
| Hawaii, U.S. | America | 60 |
| ERCOT, U.S. | America | 60 |
| Germany | Europe | 50 |
| Saudi Arabia | Asia | 60 |
| Japan | Asia | 60 |
| Northern Ireland | Europe | 50 |
| Ireland | Europe | 50 |
| England | Europe | 50 |
| Australia | Oceania | 50 |
| Bahamas | America | 60 |
| Egypt | Africa | 50 |

To promote repeatability of the simulation, the procedure of the data loading and preprocessing is outlined below and shown in Fig. 3.

Step 1: Load historical FDR measurement data files of the studied systems.

Step 2: Align the measurement data with studied periods.

Step 3: Pre-process the measurements and filter bad data.

Step 4: Normalized the measurements of the studied systems.

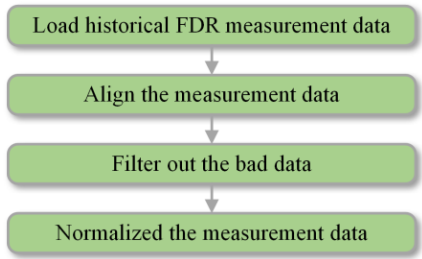

Figure 3. The procedure of FDR historical data process and analysis

### III. STATISTICAL ANALYSIS OF FREQUENCY IN POWER GRIDS

For a specific power grid, the statistic of frequency could reveal some insights into the operation status. Standard deviation and mean of the frequency measurement data are calculated in this paper. The mean of system frequency could be used to show the balance between the generation and the demand. Moreover, the standard deviation of the frequency is offered to demonstrate the frequency fluctuation. It should be noted that some operations in the power grid could also impact on the mean and standard deviation.

As shown in Fig. 4, the mean values of frequency in power grids in different regions are listed in descending order. Meanwhile, the standard deviation values of frequency are listed in descending order in Fig. 5. As shown in Fig. 5, the frequencies in all mainland power grids have smaller standard deviations, comparing to the frequencies in island power grids. One of the potential reasons for this situation is that the sizes of the mainland power grids are larger than the sizes of island power grids. However, it should be noticed that there are some exceptional cases. For Hawaii power grids, its frequency deviation is smaller than those of most mainland power grids. Another extreme case is Egypt power grid, which is located in the mainland, but its frequency deviation is the highest among the power grids studied in this paper.

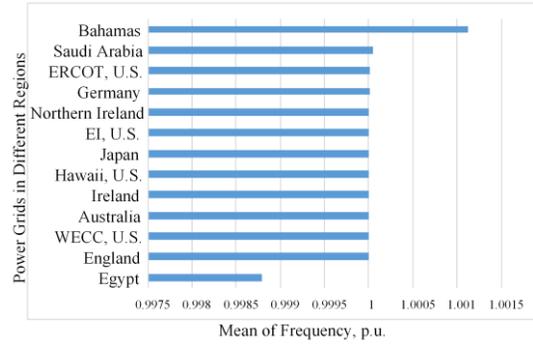

Figure 4. Mean of frequency in power grids in different regions

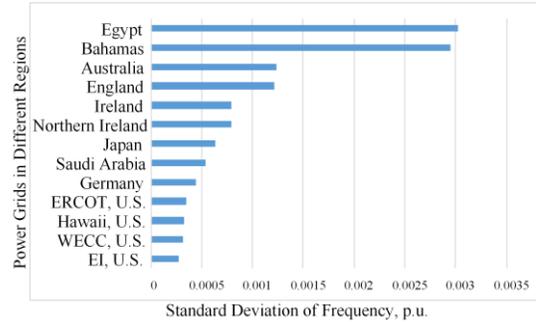

Figure 5. The standard deviation of frequency in power grids in different regions

According to [31], the highest relay setting of load shedding frequency $f$ required by North American Electric Reliability Corporation (NERC) is 59.3 Hz (0.9883333 p.u.) $<f<$59.5 Hz (0.9916667 p.u.). The measurement data includes transient status, such as generation trip and load shedding events. As shown in Fig. 5, most power grids operate with the safe frequency fluctuations in a reasonable margin, comparing to NERC load shedding frequency. However, the frequency standard deviation of Egypt power grid during the observation time in this paper is the highest, which close to the in NERC load shedding criteria.

To perform a more detailed statistical analysis, three typical power grids are selected: (1). EI system (North America, mainland power grid), (2). Egypt system (Africa, mainland power grid) and (3). Japan (Asia, island power grid). As shown in Fig. 5, there three power grids have the smallest, largest and average standard deviation in the power grids. The standard deviation and mean values of the power grids are calculated for each day. Here, the analysis time period is selected as one month. As shown in Fig. 6(a), the standard deviation of EI power grid frequency is smallest on a daily basis, compared to Japan and Egypt power grids frequency. Means of daily frequency in Egypt power grid is not a straight line as shown in Fig 6(b). It means that Egypt power grids are operated at an under-frequency status. Compared to Egypt power grids frequency mean, the means of daily frequency in EI and Japan power grids are almost flat, which indicates that the two systems are operated steadily around the nominal frequency. Also, the results indicate that the frequency fluctuation in Egypt system is higher than that in EI and Japan power grids.

In the following, the probability density function of the frequency is calculated, which is shown in Fig. 7. It can be seen that the probability density functions can divide into two types: 1) single-peak distribution, such as the probability density function of the frequency in EI, WECC, Hawaii, Germany, Japan, Australia, and Egypt; 2) multi-peak distribution, such as the probability density function of the frequency in ERCOT, Saudi Arabia, Northern Ireland, Ireland, England, and Bahamas.

As shown in Figs. 7(a)-7(g), for the single-peak distribution of the frequency, the normal distribution is considered an appropriate choice to describe it. The mean and standard deviation values of frequency are calculated and shown in Fig. 4 and Fig. 5. Based on the calculated mean and standard deviation values, the corresponding probability density functions of the frequency which almost follows normal distribution can be obtained, which is drawn as the red lines in Fig. 8. Meanwhile, the real probability density functions of the frequency statistically calculated by the observation data are also shown as the bars in Fig. 8. Obviously, the normal distribution provides a suitable profile for the single-peak distribution of the frequency.

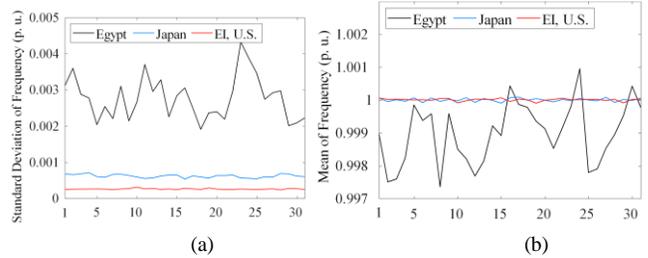

Figure 6. Standard deviation and mean of frequency in each day in three power grids. (a) Standard deviation (b) Mean.

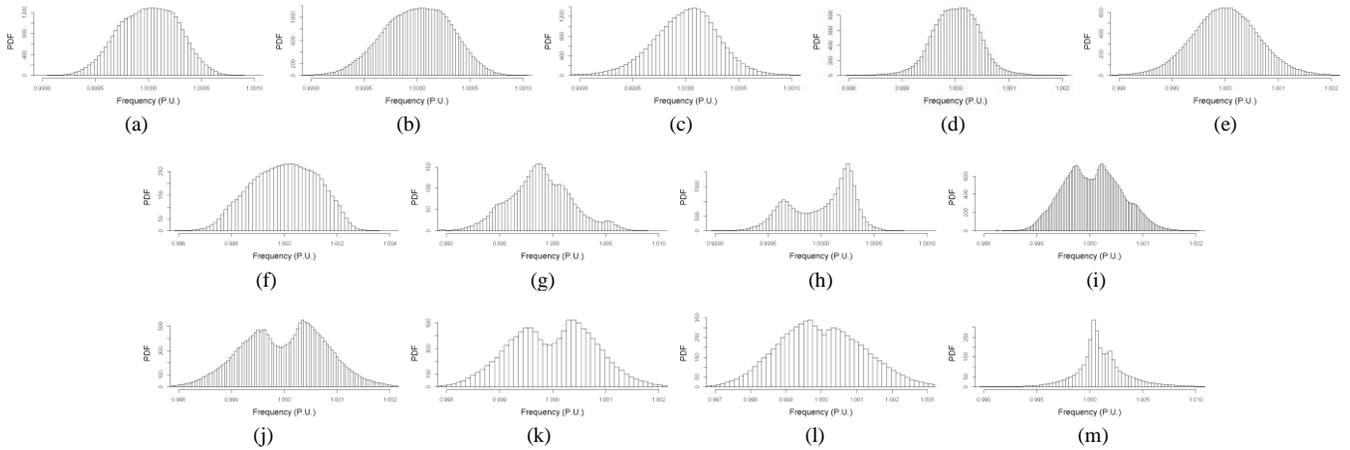

Figure 7. Probability density function of frequency in different power grids. (a) EI, U.S. (b) WECC, U.S. (c) Hawaii, U.S. (d) Germany. (e) Japan. (f) Australia. (g) Egypt. (h) ERCOT, U.S. (i) Saudi Arabia. (j) Northern Ireland. (k) Ireland. (l) England. (m) Bahamas.

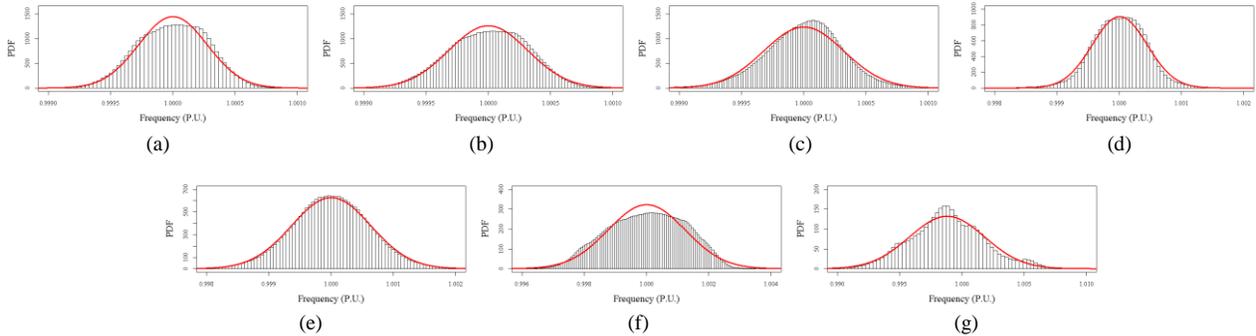

Figure 8. Comparison of probability density function obtained by statistics and the corresponding normal distribution. (a) EI, U.S. (b) WECC, U.S. (c) Hawaii, U.S. (d) Germany. (e) Japan. (f) Australia. (g) Egypt.

## IV. Conclusion

This paper utilizes the frequency measurement data provided by FNET/GridEye, to observe and statistically analyze the power frequency status of various power grids over the world, including the grids located in both mainland and islands. The comparison results show that frequencies of most power grids in mainland operate in a relatively smaller range than those in the island. From the perspective of regions, the power grids at America has the smallest frequency standard deviation. The standard deviation of Asian and European power grids frequency is at an average level in this study. The frequencies of power grids at Oceania and Africa operate at a high deviation status. Additionally, the distributions of frequency show two different categories in the worldwide power grids, i.e., the single-peak distribution and multi-peak distribution. Furthermore, a meaningful insight that the single-peak distributions of the frequency almost follow the normal distribution is found. Since the lack of the report about the frequency in the worldwide power grids, the analysis in this paper may provide some references for further research or making regulation criteria on the frequency.

## Acknowledgment

This work was supported primarily by the Engineering Research Center Program of the National Science Foundation and the Department of Energy under NSF Award Number EEC-1041877 and CURENT Industry Partnership Program.